\documentclass[a4paper,11pt]{article}
\pdfoutput=1 % if your are submitting a pdflatex (i.e. if you have
\usepackage{jheppub} % for details on the use of the package, please
\usepackage[T1]{fontenc} % if needed

\usepackage[utf8]{inputenc}
\usepackage{tikz}
\usepackage{capt-of}
\usepackage{amsmath}

\usepackage{comment}

\addtocontents{toc}{\protect\enlargethispage{\baselineskip}}

\usepackage[rightcaption]{sidecap}
\usepackage{graphicx}
\graphicspath{ {./images/} }

\newcommand{\m}{\mu}
\newcommand{\n}{\nu}

\newcommand{\vev}[1]{{\left< {#1} \right>}}

\newcommand{\cur}[1]{{\left( {#1} \right)}}
\newcommand{\squ}[1]{{\left[ {#1} \right]}}

\newcommand{\tr}{{\rm tr\,}}

\newcommand{\cB}{{\mathcal B}}

\newcommand{\cD}{{\mathcal D}}

\newcommand{\cL}{{\mathcal L}}

\newcommand{\cO}{{\mathcal O}}

\newcommand{\cG}{{\mathcal G}}

\newcommand{\cH}{{\mathcal H}}

\title{\boldmath Non-equilibrium effective field theory and second sound}

\author[a]{Michael J. Landry}

\affiliation[a]{Department of Physics, Center for Theoretical Physics,\\ Columbia University, 538W 120th Street, New York, NY, 10027, USA}

\emailAdd{ml2999@columbia.edu}

\abstract{We investigate the phenomenon of second sound in various states of matter from the perspective of non-equilibrium effective field theory (EFT). In particular, for each state of matter considered, we find that at least two (though sometimes multiple) qualitatively different EFTs exist at finite temperature such that there is always at least one EFT with a propagating second-sound wave and at least one with no such second-sound wave. To aid in the construction of these EFTs, we use the method of cosets developed for non-equilibrium systems. It turns out that the difference between the EFTs with and without second-sound modes can be understood as arising from different choices of a new kind of inverse Higgs constraint. Finally, we demonstrate that it is possible to bypass the need for new inverse Higgs constraints by formulating EFTs on a new kind of manifold  that is like the usual fluid worldvolume, but with reduced gauge symmetries.}

\arxivnumber{2008.11725}

\begin{document} 
\maketitle
\flushbottom

\section{Introduction}

Over the past decade, there has been a significant effort to understand condensed matter systems from the perspective of effective field theory (EFT). In this EFT philosophy, condensed matter systems are conceived as systems that spontaneously break spacetime symmetries. As a result, a large class of condensed matter systems can be classified in terms of their spontaneous symmetry breaking (SSB) patterns alone. Therefore, if we are only interested in the infrared (IR) behavior of such a system, the relevant degrees of freedom at zero temperature are described exclusively by Goldstone modes. It turns out that there is a very powerful technique for constructing EFTs of Goldstones known as the coset construction. This construction takes symmetries as the only input and gives an almost mechanical procedure for formulating new EFTs given a particular symmetry-breaking pattern. 

Condensed matter systems are inherently thermal and are therefore dissipative. Unfortunately, ordinary actions and Lagrangians can only give rise to conservative, that is, non-dissipative dynamics. However, recent work~\cite{Landry,Landry 2,Jensen 1,Jensen 2,H. Liu,H. Liu 2,H. Liu 3,H. Liu 2.2,H. Liu 2.3,H. Liu 2.1,H. Liu 4,FD 1,Harder,Banerjee,Jensen,Kovtun,Grozdanov,Haehl,Haehl 2,Fluid Manifesto,Hongo 1,D.V. Volkov,Hongo 2,Hongo 3,Hongo 4,Landry 3}, built upon the foundations of~\cite{Nicolis,Zoology,More gapped Goldstones,coset,Finite T superfluid}, enables the construction of effective actions that can account for dissipation and thermal fluctuations. Such actions are formulated using the in-in formalism on the Schwinger-Keldysh contour in the presence of a thermal density matrix. Further, in~\cite{Landry}, a non-equilibrium coset construction was proposed allowing the formulation of non-equilibrium EFTs for a wide range of condensed matter systems. This coset construction accounts for both ordinary Goldstones associated with SSB as well as hydrodynamic modes associated with unbroken conserved quantities at finite temperature. 

We will see, however, that EFTs of many states of matter constructed in~\cite{Landry} admit additional propagating sound modes even though such modes are not always observed in nature. These additional sound modes correspond to hydrodynamic sound waves that propagate through a fluid formed by the thermalized phonons of the system. The best-known such sound mode is the so-called second sound observed in finite-temperature superfluids, though other second-sound modes have been observed in certain crystalline solids as well~\cite{Solid second sound 1,Solid second sound 2}. It should be noted that, while second sound exists in all known superfluids, it is somewhat rare in solids. 

The aim of this paper is to use non-equilibrium EFT techniques to study the behavior of second-sound modes in various states of matter at finite temperature. Using the non-equilibrium coset construction, we will find that the presence or absence of second-sound modes derives from a new kind of inverse Higgs (IH) constraint that we can choose whether or not to impose. We investigate a wide range of condensed matter actions and find that in every case, except that of the superfluid action, the procedure to remove second sound is successful. 

We demonstrate how the whole business of imposing IH constraints to remove second-sound modes can be circumvented by defining our EFTs on a manifold other than the usual fluid worldvolume of~\cite{Landry,H. Liu,H. Liu 2,H. Liu 3}. These new worldvolumes have reduced diffeomorphism symmetries that depend on the particular state of matter in question. As an illustrative example, we construct an action for leading-order (i.e. non-dissipative) finite-temperature solids using a modified version of the non-equilibrium coset construction that is defined on a solid worldvolume. We then propose an alternative to Landau's classification of states of matter in terms of their SSB pattern. In particular, we will see that it is possible to be more precise than Landau if we instead merely specify the global and emergent gauge symmetries and make no reference to SSB at all. 

We extend the effective action for solids to leading dissipative order, finding many similarities with~\cite{Armas:2020bmo,Armas:2019sbe} and yet some disagreements as well. Finally, we explain the physical origins of the second-sound-removing IH constraints and find that they are closely connected to Umklapp scattering.

Throughout this paper we will use the ‘mostly plus convention,’ so the Minkowski metric takes
the form $\eta_{\mu\nu} =\text{diag} \cur{-,+,+,+} $.

\section{A review of relevant topics} 

Many of the concepts and mathematical techniques employed throughout this paper may be unfamiliar to some readers. Therefore, in this section, we will briefly review the main points of the zero-temperature coset construction, non-equilibrium EFT, and the non-equilibrium coset construction. In the interest of brevity, we will present many claims without proof, but we will provide references that contain more in-depth discussions of these topics.

\subsection{The zero-temperature coset construction}

Consider a Poincaré-invariant system whose full symmetry group is $\cG$ and is spontaneously broken to the subgroup $\cH$. Then, the IR dynamics are described by Goldstone modes. Since the action of the broken symmetry generators on the Goldstones is non-linearly realized, formulating an EFT for the Goldstone modes presents a challenge. Fortunately, there exists a straightforward, almost mechanical procedure for constructing the most general effective action for Goldstones. Suppose that the symmetry generators are given by
\begin{equation}\begin{split}\label{global generators} \bar P_\mu & =\text{unbroken translations},
\\ T_A & = \text{other unbroken generators},
\\ \tau_\alpha & =\text{broken generators}, 
 \end{split}\end{equation}
 where the generators $\tau_\alpha$ and $T_A$ may be some combination of internal and spacetime generators and  we have assumed that there exist some notions of spacetime translations that remain unbroken.  In this way, states can still be classified according to the corresponding notions of energy and momentum~\cite{Zoology}. Importantly, we do not require that the unbroken generators $\bar P_\mu$ be the original Poincaré translation generators (represented by $P_\m$); instead they can be some linear combination of $P_\m$ and internal symmetry generators~\cite{Zoology}. Although $\bar P_\mu$ and $T_A$ both represent unbroken generators, we will see that they play very different roles in the following construction. 
 
 Let $\cH_0$ be the subgroup of $\cH$ generated by $T_A$. 
  It turns out that to construct the most general symmetry-invariant building-blocks, it is convenient to parameterize the coset $\cG/\cH_0$ by
 \begin{equation}\label{zero T parameterization}\gamma[\pi,x) = e^{i x^\mu \bar P_\mu} e^{i\pi^\alpha(x)\tau_\alpha}, \end{equation} 
where $x^\m$ are the spacetime coordinates and $\pi^\alpha(x)$ are the Goldstone fields (up to overall normalization). Then, we may compute the Maurer-Cartan form and expand it as a linear combination of the symmetry generators by
\begin{equation} g^{-1}\partial_\m g = i E_\m^\n \cur{\bar P_\n +\nabla_\n \pi^\alpha \tau_\alpha+\cB_\n^A T_A}.  \end{equation}
It is important to note that all of the coefficients on the r.h.s. of the above expression can be explicitly computed as long as the commutators among the generators are known. It can be checked that $\nabla_\m \pi^\alpha$ is covariant under all symmetries, $\cB_\n^A$ transforms as a gauge connection and can be used to take higher-order covariant derivatives 
\begin{equation} \label{covariant derivative coset} \nabla_\m^{\cH} = (E^{-1})^\n_\m \partial_\n + i \cB_\m^A T_A ,\end{equation}
and $E_\m^\n$ plays the role of the vierbein, meaning that the invariant integration measure is $d^4 x \det E$. Then, the invariant building-blocks for the Lagrangian are formed by taking manifestly $\cH_0$-invariant combinations of the covariant objects given to us by the Maurer-Cartan form. In particular, it should be noted that the $\mu,\nu$ indices need to be contracted in ways that are invariant under the {\it unbroken} subgroup of the Lorentz symmetry group.

At leading order in the derivative expansion we have that the only covariant building-block is $\nabla_\m \pi^\alpha$; and higher-order-derivative terms are given by 
$\nabla^\cH_\m \cur{\nabla_\n \pi^\alpha},$ 
$\nabla^\cH_\m \nabla_\n^\cH \cur{\nabla_\rho \pi^\alpha},$ etc.

Finally, when only internal symmetries are spontaneously broken, the number of Goldstones equals the number of broken symmetry generators; however, when spacetime symmetries are spontaneously broken there are often fewer Goldstones than broken symmetry generators. At the level of the coset construction, we can sometimes reduce the number of Goldstones by imposing what are known as IH constraints~\cite{IH 1,IH 2}. Pragmatically, the rules of the game are as follows:
Suppose that the commutator between an unbroken translation generator $\bar P$ and a broken generator $\tau '$ contains another unbroken generator $\tau$, that is $[\bar P, \tau '] \supset \tau$. Suppose further that $\tau$ and $\tau '$ do not belong to the same irreducible multiplet under $\cH_0$. Then it turns out that it is consistent with symmetry transformations to set the covariant derivative of the $\tau$-Goldstone in the direction of $\bar P$ to zero. This equation gives a constraint that relates the $\tau '$-Goldstone to derivatives of the $\tau$-Goldstone, allowing the removal of the $\tau'$-Goldstone. The setting of this covariant derivative to zero is known as an IH constraint. 
The possible reasons for imposing these IH constraints have been investigated in~\cite{More gapped Goldstones,Low,UV completion}; they are as follows:
\begin{itemize}
\item If we were to include the Goldstones that can be removed with IH constraints, these Goldstones could appear in the effective action without any derivatives. As a result, these modes would be gapped and could therefore be integrated out. If we are only interested in the gapless degrees of freedom, then IH constraints correspond to integrating out gapped Goldstones. 
\item Sometimes when spacetime symmetries are spontaneously broken, the resulting Goldstone modes do not correspond to independent fluctuations. As a result, certain Goldstones are redundant. From this perspective, IH constraints serve as a convenient choice of `gauge-fixing' condition. 
\end{itemize}
We will see that when we apply the coset construction to non-equilibrium systems, there are more possibilities for IH constraints. 

For more on the coset construction with broken internal symmetries, consult~\cite{Weinberg} and with broken spacetime symmetries, consult~\cite{Wheel}.

\subsection{Non-equilibrium EFT}

At finite temperature, the equilibrium state is given by a mixed-stated thermal density matrix given by
\begin{equation}\rho = \frac{e^{-\beta \bar P^0}}{\text{tr} \cur{ e^{-\beta \bar P^0}}} ,\end{equation}
where $\bar P_0$ is the unbroken time-translation operator. As a result, ordinary quantum field theory  techniques that involve finding vacuum correlation functions using in-out states are of no use. Instead, we must compute quantities with the in-in formalism defined on the Schwinger-Keldysh contour~\cite{SK ref}. In this formalism, the sources are doubled. Letting $U(+\infty,-\infty;J)$ be the time-evolution operator from  the distant past to the distant future in the presence of source $J$ for some field $\Psi$, the generating functional is
\begin{equation}\begin{split}e^{W[J_1,J_2]}&\equiv\label{SK}\tr{\squ{U(+\infty,-\infty;J_1)  \rho U^\dagger(+\infty,-\infty;J_2)}} 
\\ &\equiv\int _\rho\cD\Psi_1\cD\Psi_2e^{iS[\Psi_1,J_1]-iS[\Psi_2,J_2]},\end{split}\end{equation}
where in the path integral representation, we require that in the distant future, $\Psi_1({\infty})=\Psi_2({\infty})$, and the subscript $\rho$ indicates that field configurations are weighted by the thermal density matrix functional in the infinite past.

Supposing we are only interested in the IR dynamics of this non-equilibrium system,  in typical Wilsonian fashion, we can integrate out the ultraviolet (UV) fields to obtain an effective action for the IR degrees of freedom. Let $\Psi=\{\psi^{ir},\psi^{uv}\}$, where $\psi^{ir}$ and $\psi^{uv}$ represent the IR and UV degrees of freedom, respectively. Then, we have
\begin{equation}\begin{split}e^{i I_\text{EFT}  [\psi^{ir}_1,\psi^{ir}_2;J_1,J_2]}
=\int_\rho \cD\psi^{uv}_1\cD\psi^{uv}_2 e^{i S[\psi^{uv}_1,\psi^{ir}_1;J_1]-iS[\psi^{uv}_2,\psi^{ir}_2;J_2]}.\end{split}\end{equation}
We call $I_\text{EFT}$ the non-equilibrium effective action. Notice that because it is defined on the Schwinger-Keldysh contour, the field content is doubled. 

It turns out that non-equilibrium EFTs must satisfy certain properties that can be derived from unitarity of time evolution, Wilsonian renormalization group flow arguments, and the form of the thermal density matrix. We summarize them below.

\begin{itemize}
\item The UV action describing the system of interest is factorized by $S[\Psi_1;J_1]-S[\Psi_2;J_2]. $ The effective action, however, does not admit a factorized form into the difference of two ordinary actions. In general, there exist terms that couple 1-and 2-fields in $I_\text{EFT}$. 

\item The coefficients of $S[\Psi_1;J_1]-S[\Psi_2;J_2]$ are purely real, but it turns out that the coefficients of $I_\text{EFT}[\psi^{ir}_1,J_1;\psi^{ir}_2,J_2]$ may be complex. There are three important constraints that come from unitarity, namely 
\begin{equation}\begin{split}\label{unitarity}I^*_\text{EFT}[\psi^{ir}_1,\psi^{ir}_2;J_1,J_2]&=-I_\text{EFT}[\psi^{ir}_2,\psi^{ir}_1;J_2,J_1]
\\  \text{Im} I_\text{EFT}[\psi^{ir}_1,\psi^{ir}_2;J_1,J_2]&\geq 0, ~~\text{for any} ~~\psi^{ir}_{1,2}, J_{1,2}
\\ I_\text{EFT}[\psi^{ir}_1=\psi^{ir}_2;J_1=J_2]&=0.
\end{split}\end{equation}

\item Any symmetry of the UV action $S$ is a symmetry of $I_\text{EFT}$, except for time-reversing symmetries. The fact that these time-reversing transformations are not symmetries of the effective action allows the production of entropy. Because the field values on the 1-and 2-contours must be equal in the distant future, $\psi_1^{ir}$ and $\psi_2^{ir}$ must transform simultaneously under any global symmetry transformation. Thus, there is just one copy of the global symmetry group. 

\item If the equilibrium density matrix $\rho$ takes the form of a thermal matrix, $\rho\propto e^{-\beta_0 \bar P_0}$, then the partition function $W[J_1,J_2]$ obeys what are known as the KMS conditions. These KMS conditions for the partition function can be used to derive the so-called dynamical KMS symmetries of the effective action. The way these symmetries act is as follows: Suppose that the UV theory possesses some kind of time-reversing, anti-unitary symmetry $\Theta$; at a minimum, the UV theory will be invariant under a simultaneous charge, parity, and time inversion. Then, setting the sources to zero, the dynamical KMS symmetries act on the fields by  
\begin{equation}\begin{split}\label{quantum dynamical KMS} \psi^{ir}_1(x)&\to\Theta\psi^{ir}_1(t-i\theta,\vec x),
\\ \psi^{ir}_2(x)&\to\Theta \psi^{ir}_2(t+i(\beta_0-\theta),\vec x),
\end{split}\end{equation} 
for any $\theta\in[0,\beta_0]$. It can be checked that these transformations are their own inverse, meaning that the dynamical KMS symmetries are discrete $\mathbb Z_2$ symmetries. To take the classical limit, it is convenient to perform a change of field basis by
\begin{equation}\label{r a variables} \psi^{ir}_r\equiv \frac{1}{2}\cur{\psi^{ir}_1+\psi^{ir}_2},~~~~~~~~~~\psi^{ir}_a\equiv\psi^{ir}_1-\psi^{ir}_2. \end{equation}  
Then the classical dynamical KMS symmetry transformations become 
\begin{equation}\begin{split}\psi^{ir}_r(x)&\to\Theta\psi^{ir}_r(x),
\\ \psi^{ir}_a(x)&\to\Theta\psi^{ir}_a(x)+i\Theta\squ{\beta_0\partial_t\psi^{ir}_r(x)}.
\end{split}\end{equation}
Notice that the change in $\psi^{ir}_a$ is proportional to the derivative of $\psi^{ir}_r$. Thus, when writing down terms of the effective action in the derivative expansion, it is natural to consider $\psi^{ir}_a$ and $\partial\psi^{ir}_r$ as contributing to the same order. 
\end{itemize}  

For a clearly explained, in depth review of non-equilibrium EFTs, consult~\cite{H. Liu}.

\subsection{The non-equilibrium coset construction} 

At finite temperature, SSB occurs when a symmetry generator fails to commute with the thermal density matrix. However, we may think of the thermal density matrix as an ensemble of micro-states, each of which is highly chaotic. Semi-classically, therefore, we expect each microstate to spontaneously break every symmetry of the underlying quantum field theory. As a result, in the non-equilibrium coset construction, there ought to be Goldstones associated with every symmetry generator. We will refer to Goldstones corresponding to broken generators as broken Goldstones and those corresponding to unbroken generators as unbroken Goldstones. It turns out that unbroken Goldstones enjoy a certain kind of gauge invariance, which leads to diffusion. Finally, it is most convenient to formulate non-equilibrium EFTs on the so-called `fluid world-volume' with coordinates $\phi^M$ for $M=0,1,2,3$.

Suppose that the global symmetry group is $\cG$, with generators (\ref{global generators}), and it is spontaneously broken to the subgroup $\cH$. Once again, let $\cH_0$ be the subgroup generated by $T_A$. Then, we parameterize the most general element of $\cG$ by
\begin{equation}\label{coset for non-eq} g_s(\phi) = e^{i X^\m_s(\phi)\bar P_\mu} e^{i \pi^\alpha_s(\phi)\tau_\alpha} e^{i \epsilon_s^A(\phi) T_A}, \end{equation}
where $s=1,2$ indicates on which leg of the Schwinger-Keldysh contour the fields live. Each $g_s$ for $s=1,2$ transforms under the same global symmetry action. Notice that unlike in the zero-temperature  coset construction, spacetime coordinates are now dynamical variables, $X^\m_s(\phi)$, that encode the embedding of the fluid worldvolume into the physical spacetime.  

It turns out that the non-equilibrium effective action enjoys the following gauge symmetries 
\begin{equation}\label{partial gauge}\begin{split} \phi^M & \to \phi^M + \xi^M(\phi^I),
\\ g_s(\phi)&\to g_s(\phi) e^{i \lambda^A(\phi^I) T_A},
\end{split}\end{equation} 
for arbitrary functions $\xi^M $ and $\lambda^A$ of spatial coordinates $\phi^I$ for $I=1,2,3$. 
They are `gauge' in the sense that they correspond to redundancies of description and should therefore not be thought of as physical symmetries. 

 The  Maurer-Cartan one-form is
\begin{equation}\label{MC form} g_s^{-1} \partial_M g_s =  iE_{sM}^\mu \cur{ \bar P_\mu + \nabla_{\m} \pi_s^\alpha \tau_\alpha} +i \cB^A_{sM} T_A, \end{equation}
where $E_{sM}^\mu$ are the the vierbeins, $\nabla_{\m}\pi^\alpha_s$ are the covariant derivatives of the broken Goldstones, and certain components of $\cB_{sM}^A$ behave like gauge connections.
The building-blocks that transform covariantly under both the global symmetries and the gauge symmetries~(\ref{partial gauge}) are as follows: First, there are the building-blocks from the usual coset construction, namely $\nabla_{\m } \pi^\alpha_s$, which transform covariantly under~(\ref{partial gauge}),  and  to take higher-order covariant derivatives we can use 
\begin{equation}\label{cov. deriv.ative} \frac{\partial}{\partial \phi^0},~~~~~~~~~~\nabla^\cH _{I} = {\partial_I + i \cB_{rI}^A T_A}, \end{equation}
where $\cB_{rI}^A =\frac{1}{2}\cur{\cB_{1I}^A+\cB_{2I}^A} $ and $I=1,2,3$. 
 To contract coordinate indices, we use the metrics
\begin{equation}G_{sMN} =  E^\m_{sM} \eta_{\m\n} E^\n_{sN}. \end{equation} 
Second, there are new building-blocks that involve the unbroken Goldstone degrees of freedom, 
namely $E_{s0}^\mu$ and $\cB_{s0}^A$, which transform covariantly. Finally, we have terms that involve combinations of 1-and 2-fields. Notice that $E_{1M}^\m (E_2^{-1})^M_\n$ and $\cB_{aM}^A\equiv \cB_{1M}^A-\cB_{2M}^A$ transform covariantly and we can contract coordinate indices with $E^\m_{1M}\eta_{\m\n}E^\n_{2N}$.

Often we may impose IH constraints to remove extraneous Goldstone modes. In addition to the ordinary IH constraints that also exist in the zero-temperature case, there are two new IH constraints that only exist at finite temperature. They are as follows:
\begin{itemize}
\item {\it Thermal IH:} Suppose that at finite temperature, the commutator between a broken generator $\tau$ and the unbroken time-translation generator $\bar P_0$ contains an unbroken spacetime translation generator $\bar P$, that is $[\tau,\bar P_0]\supset \bar P$. Then we may set to zero the component of $E_0^\m$ in the direction of $\bar P$. This equation can be solved algebraically to yield an expression for the $\tau$-Goldstone in terms of derivatives of the $\bar P$-Goldstone. This expression allows the removal of the $\tau$-Goldstone. 
\item {\it Unbroken IH:} Suppose that at finite temperature, the commutator between an unbroken generator $T$ and an unbroken spacetime translation generator $\bar P'$ contains another unbroken spacetime translation generator $\bar P $, that is $[T,\bar P']\supset \bar P$. Consider the matrix ${A^\m}_\n\equiv  (E_1)_M^\m (E_2^{-1})^M_\n$, where $M=0,1,2,3$ are coordinate indices and $\m,\n=0,1,2,3$ are Lorentz indices. Then we may set to zero the components of $A_{\m\n}$ in the directions of $\bar P$ and $\bar P' $. Suppose that under the dynamical KMS symmetry transformation, $A_{\m\n}\to \tilde A_{\m\n}$. Then, we may also set to zero the components of $\tilde A_{\m\n}$ in the directions of $\bar P$ and $\bar P' $. These conditions give constraints that relate the $T$-Goldstones to derivatives of the $\bar P$-Goldstones, allowing the removal of the $T$-Goldstones.
\end{itemize}

We will see that there is a new kind of IH constraint responsible for removing second-sound modes, however unlike the other IH constraints it is not derivable from purely algebraic considerations. 

Finally, we must impose the dynamical KMS symmetries. For many of the examples that we will consider, at leading order in derivatives, the only effect of these symmetries is to force the effective action to factorize into the difference of two ordinary actions, that is
\begin{equation} I_\text{EFT}[\psi_1^{ir},\psi_2^{ir}] = S[\psi_1^{ir}]-S[\psi_2^{ir}] + \cdots. \end{equation}
For simplicity, we will almost exclusively work to leading order in the derivative expansion, in which case we will deal with just one copy of the ordinary action. However, our treatment of dissipative solids and liquid crystals and the explanation of the origin of the second-sound-removing IH constraints will require doubled field content. For these actions, the dynamical KMS symmetries will be important. 

For more on the non-equilibrium coset construction, consult~\cite{Landry}. 

\section{IH constraints and second sound}

Before we proceed to computing particular non-equilibrium effective actions, it is important to understand the meaning of second sound and how it might relate to IH constraints. In many states of matter, some spatial translations are spontaneously broken. Suppose that $\bar P_{\bar\m}$ for a particular, fixed value of $\bar \mu=0,1,2,3$  is spontaneously broken, but that there exists some internal $U(1)$ symmetry generator $Q_{\bar \m}$ such that $\bar P_{\bar \m}\equiv P_{\bar \m}+Q_{\bar \m}$ remains unbroken. Then, the unbroken Goldstone modes on the Schwinger-Keldysh contour corresponding to $\bar P_{\bar \m}$ are $X^{\bar \m}_s(\phi) $ and we denote the the broken Goldstone modes corresponding to $Q$  by $\pi_s^{\bar \m}$, where $s=1,2$ indicates on which leg of the Schwinger-Keldysh contour the fields are defined. Thus, we have two kinds of Goldstone modes corresponding to translations along the $\bar \mu$ direction, meaning that we have two different kinds of sound-modes. 

After computing the Maurer-Cartan form, we find that the covariant building-blocks include the terms
${\partial \psi_s^{\bar \m}}/{\partial \phi^0}$,
where $\psi_s^{\bar \m}\equiv  X^{\bar \m }+\pi^{\bar \m}$. It is therefore consistent with symmetries to fix
\begin{equation}\label{generic IH} \frac{\partial \psi_s^{\bar \m}}{\partial \phi^0}=\delta^{\bar \m}_0.\end{equation}
Transforming to $r$-and $a$-type variables (\ref{r a variables}), the above equations are insufficient to remove $X_a^{\bar\m}$; however, it is consistent with symmetries, including dynamical KMS, to fix
\begin{equation}\label{generic IH a type} \psi_a^{\bar\m}=0\implies X_a^{\bar \m} = - \pi_a^{\bar \m},~~~~~~~\bar\mu\neq 0,\end{equation}
thereby removing $X_a^{\bar \m}$ entirely. We will see later on that there are significant problems with the $\bar\mu=0$ IH constraint. Next, by inverting $X_r^{ \mu}(\phi) $ and taking the classical limit, we can define our effective action on the physical spacetime coordinates $x^\m$ such that $\phi^M(x)$ for $M=0,1,2,3$ are now the $r$-type $\bar P_{\bar \m}$-Goldstones~\cite{H. Liu}. It turns out that (\ref{generic IH}) is sufficient to remove $\phi^{\bar \m}$ as an independent degree of freedom; we will see how this can be done in the following sections. 
 
Using this procedure, we thus successfully remove the unbroken Goldstone mode corresponding to $\bar P_{\bar \m}$. We claim that imposing all such possible IH constraints kills the second-sound mode; we will see explicitly that this is the case in the following examples. Finally notice that the conserved current associated with $Q_{\bar \m}$ denoted by ${J_{\bar \m}}^\m$ and the $\bar\mu$-component of the stress energy tensor, ${T_{\bar\m}} ^\m $ are now identified with one another, up to an overall minus sign. 
 
Throughout the following sections, we will repeatedly encounter many of the same building-blocks. Defined on the fluid worldvolume, they are
\begin{equation}\begin{split} \label{generic building-blocks}
\\ G_{MN} & = \frac{\partial X^\m}{\partial\phi^M}\eta_{\m\n}\frac{\partial X^\n}{\partial\phi^N},
\\ Y^{\m\n} & = G^{MN} \partial_M\psi^\m\partial_N\psi^\n,
\\ Z^{\m} & = \frac{\partial \psi ^\m}{\partial\phi^0},
\end{split}\end{equation} 
where $G^{MN}$ is the inverse of the pull-back metric $G_{MN}$. Transforming to the physical spacetime, we find that 
\begin{equation}\begin{split}
({-G_{00}})^{-1/2} & \to \tau\equiv u^\m \partial_\m \phi^0,
\\ Y^{\m\n}&\to y^{\m\n}\equiv \partial_\rho \psi^\m\partial^\rho\psi^\n,
\\ Z^\m &\to \zeta^\m \equiv \frac{1}{\tau} u^\m \partial_\m\psi^\m,
\end{split}\end{equation}
where $u^\mu= J^\mu/\sqrt{-J^2}$ such that $\star J = d\phi^1\wedge d\phi^2 \wedge d\phi^3$.

\section{Superfluids}

As a warm-up, we will demonstrate how to remove second sound from superfluids. While no such superfluids have been observed in nature, this toy model provides a simple example to see how our new IH constraint can be imposed. It is important to note, however, that imposing the IH constraint of the form~\eqref{generic IH a type} is necessary to remove $a$-type fields. Unfortunately, for superfluids we are interested in the $\bar\mu=0$ case, which is not consistent with dynamical KMS symmetries. However, if we focus only on the leading-order action, we can work with just one copy of the fields, thereby (mathematically) circumventing the problem. 
We will comment on how these unusual superfluids without second sound might be realized in nature and why they ordinarily are not in \S\ref{IH meaning}.

Consider a finite-temperature superfluid. Since our theory is relativistic, it ought to be Poincaré-invariant. In our `mostly plus' convention, the Poincaré algebra is 
\begin{equation}\begin{split}\label{Poincare}
i [J_{\m\n},J_{\rho\sigma}] &= \eta_{\n\rho} J_{\m\sigma} -\eta_{\m\rho}J_{\n\sigma} -\eta_{\sigma\m} J_{\rho\n} +\eta_{\sigma\n} J_{\rho\m},
\\ i[P_\m, J_{\rho\sigma}] & = \eta_{\m\rho} P_\sigma-\eta_{\mu\sigma} P_\rho,
\\ i[P_\m,P_\n] & = 0, 
\end{split}\end{equation}
where $P_\m$ are the translation generators and $J_{\m\n}$ are the Lorentz generators. 
From the EFT perspective, a superfluid is defined as a system that has a conserved $U(1)$ charge $Q$ such that both $Q$ and $P_0$ (i.e. time translations) are spontaneously broken but a diagonal subgroup, $\bar P_0\equiv P_0+\mu_0 Q$ is preserved\footnote{We include the factor of the equilibrium chemical potential $\mu_0$ as a matter of convention.}~\cite{Witten,Son}. As a result, the broken generators are $Q$, corresponding to conserved particle number and $K_i\equiv J_{0i}$, corresponding to Lorentz boosts. The unbroken translations are $\bar P_0$ and $P_i$, and the remaining unbroken generators are $J_i =\frac{1}{2} \epsilon^{ijk} J_{jk}$, corresponding to spatial rotations. The most general group element is 
\begin{equation} g(\phi) = e^{i X^\m(\phi)\bar P_\mu} e^{i\pi(\phi) Q} e^{i\eta^i(\phi) K_i} e^{i\theta^i(\phi) J_i}.\end{equation}
Following the steps of~\cite{Landry}, and converting to physical spacetime, we find that the leading-order action is 
\begin{equation}\label{coset finite T superfluid} S_\text{S.S.} =\int d^4x ~P(\tau,y^{00}, \zeta^0),  \end{equation} 
where the subscript S.S. stands for `second sound.' 

To see that this action has a second-sound mode, let us expand the action to quadratic order in small fluctuations. Letting $\phi^\m=x^\m +\varepsilon^\m(x)$ and performing suitable field-redefinitions to decouple $\varepsilon$ and $\pi$, we have that the quadratic Lagrangian takes the form
\begin{equation}\begin{split} \label{quadratic superfluid 1} \cL^{(2)} = \frac{1}{2}\Big[ C_0 (\dot\varepsilon^0)^2 + C_1 (2\dot\varepsilon^i \partial_i \varepsilon^0+(\dot\varepsilon^i)^2) 
+ M_0\dot \pi^2-M_1(\partial_i \pi)^2\Big].  \end{split}\end{equation} 
It is then straight-forward to check that there are two wave-solutions, corresponding to first and second-sound modes. The first sound mode arises form the superfluid degrees of freedom $\pi$. The corresponding speed of sound squared is $c_1^2 = M_1/M_0$. The second sound mode corresponds to waves in the ordinary fluid degrees of freedom, $\varepsilon^\m$. The corresponding speed of sound squared is $c_2^2 = C_1/C_0$. 

Now impose the IH constraint 
\begin{equation}\label{superfluid IH}\frac{\partial \psi}{\partial \phi^0} = 1. \end{equation} 
Transforming to physical spacetime, this constraint yields $\zeta^0=1$, which can be rearranged to give
\begin{equation} \tau\equiv  u^\m \partial_\m\phi^0=u^\m\partial_\m \psi . \end{equation}
Thus, since $\phi^0$ only appears in the action through the building-block $\tau$, we have successfully removed it as an independent degree of freedom. In particular, anywhere $\phi^0$ appears, we may replace it with $\psi$. Thus, the building-blocks are $y^{00}$, which is unaffected by the constraint (\ref{superfluid IH}), and $\tau_0\equiv u^\m\partial_\m \psi$. The resulting effective action is
\begin{equation}\label{coset finite T superfluid} S_\text{no S.S.} =\int d^4x ~P(\tau_0,y^{00}).  \end{equation} 

Now consider the quadratic action. The IH constraint (\ref{superfluid IH}), at the linearized level, gives $\dot\varepsilon^0 = \dot \pi$. Thus, the quadratic action becomes
\begin{equation}\begin{split}\label{quadratic superfluid 2} \cL^{(2)} = \frac{1}{2}\Big[ (C_0+M_0)\dot\pi^2 + C_1 (2\dot\varepsilon^i \partial_i \pi+(\dot\varepsilon^i)^2) -M_1(\partial_i \pi)^2\Big].  \end{split}\end{equation} 
It is straightforward to check that there is now only one propagating sound wave solution with speed of sound squared $c_s^2 = C_1/(C_0+M_0)$. Thus, we have an EFT for superfluids at finite temperature with just one sound mode. %Physically, this corresponds to the ultra low-temperature regime of superfluids in which the thermal excitations are too small to form collective-mode second-sound waves. 

Finally, we investigate the local thermodynamic behavior of the system. Letting $T_0$ and $\mu_0$ be the equilibrium temperature and chemical potential, respectively, the local temperature is $T\equiv T_0\tau$ and the local chemical potential is $\mu\equiv \mu_0 \tau_0$~\cite{Landry, H. Liu}. We see that before imposing the IH constraint (\ref{superfluid IH}), the local temperature and chemical potential can fluctuate independently; however after imposing (\ref{superfluid IH}), they are proportional, namely
\begin{equation}\frac{T}{T_0} = \frac{\mu}{\mu_0}. \end{equation}

\section{Solids} 

We now turn our attention to finite-temperature solids. We will find that, like in the superfluid case, the second-sound mode of solids can be removed with IH constraints. In the case of isotropic solids, there are just two types: those with second sound and those without. However in anisotropic solids, it is possible to impose anywhere between zero and three IH constraints. If we impose some but not all possible constraints, then we may have a second-sound mode that can propagate in some directions, but not others. 

Solids spontaneously break spatial translations and rotations, but to ensure that some sort of unbroken momentum exists, we must introduce three broken internal $U(1)$ symmetry generators $Q_i$ for $i=1,2,3$ such that $\bar P_i\equiv P_i+Q_i$ remain unbroken. Thus the broken generators are $Q_i$, $J_i$, and $K_i$ and the unbroken generators are $\bar P_\m$, where $\bar P_0\equiv P_0$.  The most general group element is
\begin{equation} g(\phi)=e^{i X^\mu(\phi)\bar P_\mu} e^{i\pi^i(\phi)Q_i} e^{i\eta^i(\phi)K_i}e^{i\theta^i(\phi)J_i}. \end{equation}
Imposing the IH constraints of~\cite{Landry} and transforming to physical spacetime, we find that the leading-order action is
\begin{equation}\label{solid Lagrangian 1} S_\text{S.S.} = \int d^4x~P(\tau,y^{ij},\zeta^i).  \end{equation}

\subsection{Isotropic solids}

To simplify the problem, suppose that the solid is isotropic; then we must introduce an internal $SO(3)$ symmetry with generators $S_i$ for $i=1,2,3$ such that $J_i+S_i$ is unbroken. Moreover we require that $[S_i,Q_j]=i\epsilon_{ijk} Q_k$. Rather than going through the coset construction again with these additional symmetries, we merely comment that the corresponding Goldstones can be removed with IH constraints. As a result, the only effect of the unbroken rotational symmetry is that all spatial indices $i,j,\dots$ must be contracted in manifestly $SO(3)$-invariant ways. Then, the Lagrangian (\ref{solid Lagrangian 1}) can only depend on $y^{ij}$ and $\zeta^i$ in the combinations 
\begin{equation}\label{iso combinations} \tr y,~~~~~~~~~~\tr y^2,~~~~~~~~~~\tr y^3\end{equation}
and
\begin{equation}\label{iso combinations 2}(\zeta^i)^2, ~~~~~~~~~~ y^{ij}\zeta^i \zeta^j,~~~~~~~~~~ (y^{ij} \zeta^j)^2.\end{equation}
Expanding the isotropic solid Lagrangian to quadratic order in the fields and making suitable variable changes to decouple the fluid and solid degrees of freedom, we find that 
\begin{equation}\begin{split} \label{quadratic iso solid 1} \cL^{(2)} = \frac{1}{2}\Big[ C_0 (\dot\varepsilon^0)^2 + C_1 (2\dot\varepsilon^i \partial_i \varepsilon^0+(\dot\varepsilon^i)^2) 
+M_0 (\dot \pi^i)^2-M_1(\partial_i \pi^i)^2 - M_2 (\epsilon^{ijk} \partial^j \pi^k)^2\Big].  \end{split}\end{equation} 
It is straightforward to check that the sound waves can be classified as follows: a longitudinal solid mode with speed squared $c_L^2 = M_1/M_0$, a transverse solid mode with speed squared $c_T^2 = M_2/M_0$, and a longitudinal hydrodynamic second sound with speed squared $c_2^2 = C_1/C_2$. 

Let us impose the additional IH constraints; there are three of them, namely
\begin{equation}\label{iso solid IH con}\frac{\partial \psi^i}{\partial\phi^0} = 0. \end{equation}
Notice that because of isotropy, we must impose all three conditions simultaneously. We will see in the next subsection that if isotropy is broken, we have more options. Converting the above equation to physical spacetime, we have
\begin{equation} \label{iso solid ih}u^\m \partial_\m \psi^i=0. \end{equation}
Notice that the symmetries (\ref{partial gauge}) require that $\phi^I$ may appear in the effective action only in the package $u^\m$. But the above constraints require that $u^\m$ be orthogonal to the vectors $\partial_\m \psi^i$ for $i=1,2,3$. With the assumption that $u^\m$ remain orthochronous, we find that (\ref{iso solid ih}) implies $u^\m=v^\m$ such that
\begin{equation} v^\m \equiv \frac{j^\m}{\sqrt{-j^2}},~~~\star j = d\psi^1\wedge d\psi^2\wedge d\psi^3. \end{equation}
Thus, we have successfully removed the unbroken Goldstones $\phi^I$ for $I=1,2,3$. The covariant building-blocks for the effective action are now $\tau_1\equiv v^\m \partial_\m\phi^0$ and $y^{ij}$, so we have 
\begin{equation}\label{solid Lagrangian 2} S_\text{no S.S.} = \int d^4x~P(\tau_1,y^{ij}),  \end{equation}
where it is understood that $y^{ij}$ appears only in the forms given by (\ref{iso combinations}). 

 At the level of the quadratic action, imposing these IH constraints gives us the linearized relations $\dot \epsilon^i =\dot \pi^i$. Thus, the quadratic Lagrangian becomes
\begin{equation}\begin{split} \label{quadratic iso solid 2} \cL^{(2)} = \frac{1}{2}\Big[ C_0 (\dot\varepsilon^0)^2 +2  C_1 \dot\pi^i \partial_i \varepsilon^0+
(C_1+M_0) (\dot \pi^i)^2 -M_1(\partial_i \pi^i)^2 - M_2 (\epsilon^{ijk} \partial^j \pi^k)^2\Big].  \end{split}\end{equation} 
We now have one transverse solid sound wave with speed squared $c_T^2 = M_2/(C1+M_0)$ and one longitudinal sound wave with speed squared $c_L^2 = ( M_1+C_1^2/C_1)/(C_1+M_1)$. Thus, there is no longer a fluid-like second-sound wave; this is typical of most solids.

\subsection{Uniaxial crystals}

Now, we investigate the simplest case of an anisotropic solid, namely the uniaxial crystal. We will take the $\hat z$-direction to be the axis of symmetry and let $A,B=1,2$ label the directions perpendicular to the symmetry axis. Then this crystal has an internal $SO(2)$ symmetry generated by $S_3$ such that $J_3+S_3$ remains unbroken. The effect of this symmetry will be to force $A,B$ indices to contract in manifestly $SO(2)$-invariant ways. Thus $y^{ij}$ and $\zeta^i$, may appear in the effective action only in the packages~\eqref{iso combinations}, \eqref{iso combinations 2} and 
\begin{equation} y^{33}, ~~~~~ (y^{3A})^2,~~~~~\zeta^3,~~~~~y^{3A} \zeta^A. \end{equation}
Performing the field redefinitions necessary to decouple $\epsilon^\m$ from $\pi^i$, the leading-order quadratic Lagrangian is
\begin{equation}\begin{split} \label{quadratic uniaxial 1} \cL^{(2)} = \frac{1}{2}\Big[ C_0 (\dot\varepsilon^0)^2 + C_1 (2\dot\varepsilon^A \partial_A \varepsilon^0+(\dot\varepsilon^A)^2)
 + C_3 (2\dot\varepsilon^3 \partial_3 \varepsilon^0+(\dot\varepsilon^3)^2)+M_0 (\dot \pi^i)^2
\\ -M_{3L}(\partial_3 \pi^3)^2 - M_{3T}  (\partial^A \pi^3)^2
- M_{1L} (\partial_A\pi^A)^2 - M_{1T} (\partial_3 \pi^A)^2 
\\- M_{2T} (\epsilon^{3AB}\partial_A\pi^B)^2
-M_{4}(\partial_3\pi^3 \partial_A\pi^A)\Big].  \end{split}\end{equation} 
The dispersion relations of the solid degrees of freedom are rather complicated, but it can be checked that they agree with the usual dispersion relations of uniaxial crystals. Further, it is easy to see that the fluid degrees of freedom $\epsilon^\m$ have a longitudinal second-sound wave solution with sound speed squared $c_2^2=C_1/C_0$ when it propagates in the $x$-$y$ plane and $c_3^2=C_3/C_0$ when it propagates parallel to the $z$ axis. 

Now, we could impose the IH constraints~(\ref{iso solid IH con}) as we did before, but this will not give us anything too new. Instead, we will exploit the anisotropy of the uniaxial crystal. We have two anisotropic options for IH constraints, namely
\begin{equation} \frac{\partial\psi^A}{\partial\phi^0} = 0~~~\text{OR}~~~ \frac{\partial\psi^3}{\partial\phi^0} = 0.\end{equation}
If we impose both sets of constraints simultaneously, the second-sound mode is killed entirely; however, if we impose just one, then the second sound is not entirely removed, though its dynamics are restricted. 
In particular, imposing the first set of constraints prevents the second-sound wave from propagating in the $x$-$y$ plane, while imposing the second prevents the second-sound wave from propagating parallel to the $z$ axis.

\subsection{Supersolids}

Supersolids are just like ordinary solids except that now $P_0$ is spontaneously broken and there exists a $U(1)$ charge $Q_0$ such that $\bar P_0\equiv P_0+Q_0$ remains unbroken. For simplicity, assume the supersolid is isotropic. Then, the effective action is given by
\begin{equation} S_\text{S.S.} =  \int d^4 x ~P(\tau, y^{00},y^{0i},y^{ij},\zeta^0,\zeta^i),\end{equation} 
where it is understood that $y^{0i}$, $y^{ij}$, and $\zeta^i$ are contracted in $SO(3)$-invariant ways. We now have two possible sets of IH constraints\footnote{If we impose the $\bar\mu=0$ IH constraint from~\eqref{generic IH a type}, we will encounter the same problems that plague the superfluid. But in the leading-order action, we can (mathematically) get away with ignoring these issues.} 
\begin{equation} \frac{\partial\psi^0}{\partial\phi^0}=0~~~\text{OR}~~~\frac{\partial\psi^i}{\partial\phi^0}=0. \end{equation}
We can impose zero, one, or both of these. If we impose the first, then just as in the superfluid case, the second-sound mode is killed; if we impose the second, then just as in the isotropic solid case, the second-sound mode is killed. If we impose both sets of IH constraints, then the leading-order action at finite temperature is 
\begin{equation} S_\text{S.S.} =  \int d^4 x ~P(y^{00},y^{0i},y^{ij}),\end{equation} 
which is identical to the leading-order action at zero temperature~\cite{Zoology,coset}.

\section{Smectic liquid crystals}

Liquid crystals are states of matter that exist on a spectrum somewhere between fluids and crystalline solids. Crystalline solids spontaneously break all spatial translations and rotations, but preserve a discrete subgroup of translations, whereas fluids do not break any translations or rotations. Smectic liquid crystals consist of stacked layers of molecules; in this way spatial translations along one direction are spontaneously broken~\cite{unified hydro,Smectic A paper}. Without loss of generality, we will take this broken translation generator to be $P_3$. Additionally, the presence of the sacked layers breaks the rotations $J_1$ and $J_2$. To ensure that some notion of translations is preserved, we must introduce a $U(1)$ charge $Q_3$ that is spontaneously broken such that the diagonal subgroup generated by $\bar P_3\equiv P_3+Q_3$ is preserved. We will use indices $A,B=1,2$ to indicate directions orthogonal to the stacked layers. 

\subsection{Phase A}

In phase A, translations in the $x$-$y$ plane and rotations about the $z$ axis are unbroken. Thus, $ \bar P_\m = P_\m+\delta^3_\m Q_3$, and $J_3$
are the unbroken generators and
$ J_A$, $K_i$, and $Q_3$
are the broken generators. The most general group element is
\begin{equation}\begin{split}  g(\phi) = e^{i X^\m (\phi)\bar P_\m} e^{i \pi^3(\phi) Q_3}  e^{i\eta^3(\phi) K_3+i \theta^{A}(\phi) J_{A}+{i} \theta^{3}(\phi) J_{3}} e^{i \eta^A(\phi) K_A} .  \end{split}\end{equation} 
Going through the steps given in~\cite{Landry}, and transforming to physical spacetime, we find the leading-order action is
\begin{equation} S_{S.S.} =\int d^4 x~ P(\tau,y^{33},\zeta^3).  \end{equation}
Expanding to quadratic order in the fields and performing the necessary field redefinitions to decouple $\epsilon^\m$ from $\pi^3$, we arrive at the quadratic Lagrangian
\begin{equation}\begin{split} \label{quadratic phase A 1} \cL^{(2)} = \frac{1}{2}\Big[ C_0 (\dot\varepsilon^0)^2 + C_1 (2\dot\varepsilon^A \partial_i \varepsilon^0+(\dot\varepsilon^A)^2) 
+ C_3 (2\dot\varepsilon^3 \partial_i \varepsilon^0+(\dot\varepsilon^3)^2) + M_0 (\dot \pi^3)^2 \\- M_1 (\partial_A \pi^3)^2 -M_3(\partial_3 \pi^3)^2\Big].  \end{split}\end{equation} 
Notice that there is one transverse and one longitudinal mode that propagate in the $x$-$y$ plane with speeds of sound squared $c_T^2= M_1/M_0$ and $c_L^2=C_1/C_0$, respectively. And there are two longitudinal modes that propagate parallel to the $z$ axis with speeds of sound squared $c_{3L}^2= M_3/M_0$ and $c_2^2=C_3/C_0$. Thus, there are two longitudinal waves that can propagate in the same direction, meaning that our theory supports a second-sound mode. 

Now impose the IH constraint
\begin{equation} \frac{\partial\psi^3}{\partial \phi^0} = 0. \end{equation} 
Converting to physical spacetime, this give 
\begin{equation} u^\m\partial_\m \psi^3 = 0.\end{equation}
Thus, given that $u^\m$ is orthochronous and orthogonal to $\partial_\m \phi^A$ for $A=1,2$, the above equation means that we may replace all instances of $u^\m$ by
\begin{equation} u^\m \to v^\m_3 \equiv \frac{j^\m_3}{\sqrt{-j_3^\m j_{3\m}}},~~~~~\star j_3 \equiv d\phi^1\wedge d\phi^2 \wedge d\psi^3. \end{equation}
Therefore, all instances of $\phi^3$ in the effective action can be replaced with $\psi^3$. As a result, the building-blocks are $\tau_3\equiv v_3^\m\partial_\m\phi^0$ and $y^{33}$, so we have
\begin{equation}S_\text{no S.S.} = \int d^4 x~ P(\tau_3,y^{33}). \end{equation}
At the linearized level, these IH constraints become $\dot \epsilon^3 = \dot \pi^3 $. The resulting quadratic action~is 
\begin{equation}\begin{split} \label{quadratic phase A 2} \cL^{(2)} = \frac{1}{2}\Big[ C_0 (\dot\varepsilon^0)^2 + C_1 (2\dot\varepsilon^A \partial_i \varepsilon^0+(\dot\varepsilon^A)^2) 
+ 2 C_3 \dot\pi^3 \partial_i \varepsilon^0 + (M_0+C_3) (\dot \pi^3)^2 \\- M_1 (\partial_A \pi^3)^2 -M_3(\partial_3 \pi^3)^2\Big].  \end{split}\end{equation} 
We see therefore that there is now just one longitudinal sound wave propagating parallel to the $z$ axis, meaning that we have successfully removed the second-sound mode. Notice that even though there is no second-sound mode, we still have a hydrodynamic sound mode that can propagate in the $x$-$y$ plane; however this mode is not a {\it second} sound mode as there are no other such longitudinal sound waves that can propagate in the same direction. 

\subsection{Phase B}

Phase B smectic liquid crystals are essentially just solids that cannot sustain uniform $x$-$z$ or $y$-$z$ shears~\cite{unified hydro}. At the level of effective field theory, this inability to sustain such shears is captured by the symmetries
\begin{equation}\label{phase B symmetries} \psi^A\to\psi^A+g^A(\psi^3), \end{equation} 
for arbitrary functions $g^A$~\cite{Landry}. Thus, the effective action is just~(\ref{solid Lagrangian 1}) except $y^{ij}$ can only appear in the packages
\begin{equation}\begin{split} b\equiv \det y^{ij}, ~~~&~~~y^{33},
\\ b_1\equiv y^{11}y^{33}-(y^{13})^2,~~~&~~~ b_2\equiv y^{22}y^{33}-(y^{23})^2, \end{split}  \end{equation}
and $\zeta^A$ for $A=1,2$ cannot appear in the effective action. 

Imposing the IH constraints
\begin{equation} \frac{\partial\psi^i}{\partial\phi^0}=0, \end{equation}
just as in the solid case, allow us to replace all instances of $\phi^i$ with $\psi^i$. And again, just as in the solid case, these IH constraints remove the second-sound wave.

Since smectic phase B is anisotropic one might wonder if we can impose anisotropic IH constraints. If we were dealing with an ordinary anisotropic solid, we could freely mix and match from the constraints
\begin{equation}\frac{\partial\psi^1}{\partial\phi^0}=0~~~\text{OR}~~~\frac{\partial\psi^2}{\partial\phi^0}=0~~~\text{OR}~~~\frac{\partial\psi^3}{\partial\phi^0}=0.\end{equation}
However, because of the additional symmetries~(\ref{phase B symmetries}), imposing either of the first two constraints without also imposing the third is prohibited. Thus, if we are to impose any IH constraints at all, we must impose ${\partial\psi^3}/{\partial\phi^0}=0$, which will prevent second-sound waves from propagating parallel to the $z$ axis.

\section{The meaning of IH constraints}\label{IH meaning} 

It is a curious fact of nature that second sound always exists in superfluids but not in other states of matter like solids. Why should this be the case? To understand why, suppose that $\bar P_{\bar \m} = P_{\bar\mu}+Q_{\bar\mu}$. By imposing the constraints 
\begin{equation}\label{IH stuff.}\frac{\partial \psi_r^{\bar\mu}} {\partial\phi^0} = \delta_0^{\bar\mu}, ~~~~\psi_a^{\bar\mu}=i \beta_0 \delta^{\bar \mu}_0,\end{equation}
we are essentially removing $Q_{\bar \mu}$ as an independently conserved quantity. Thus, while the mathematical possibility of superfluids with no second sound exists, it requires the non-conservation of the $U(1)$ charge $Q$ associated with particle number. Removing the conservation of $Q$ does not make sense as particle number conservation has physical meaning independent of superfluid phase.  For solids, however, the internal translation generators $Q_i$ emerge from the periodicity of the solid lattice and hence are only defined in solid phase. As soon as the solid melts, these symmetries simply vanish. From this perspective, it is not so strange that an IH constraint should be able to remove them entirely from the physical theory. Viewed from another perspective, it is well-known that second sound only exists in solids with a pristine crystalline lattice structure and low probability of Umklapp scattering~\cite{solid second sound ref}. The reason is that Umklapp scattering leads to non-conservation of lattice momentum; i.e. $Q_i$ is not conserved by Umklapp scattering events. Since superfluids have no lattice structure, there can never be Umklapp scattering and hence second sound must always persist. There has, however, been recent work indicating that systems with periodic structure in time can exist in thermodynamic equilibrium states, known as time crystals~\cite{time crystal}. This leaves open the intriguing possibility that the superfluids or supersolids without second sound may describe time crystals in the limit of large Umklapp scattering. There are, however, reasons to doubt this interpretation. We will discuss them at the end of this section. 

To see explicitly how Umklapp scattering can lead to IH constraints, we now consider the simple example of a solid in the limit of large Umklapp scattering. We begin by postulating the existence of a second-sound mode and show that when Umklapp scattering is large, integrating out the Goldstones $\psi^i$ associated with the internal translation generators $Q_i$ is equivalent to imposing the usual IH constraints. At leading order in the derivative expansion, the ordinary effective action for a solid with second sound is given by~(\ref{solid Lagrangian 1}). For this exercise, however, we wish to work with the non-equilibrium EFT with doubled field content. Woking to leading order in the derivative expansion and using the $r,a$-basis, we have
\begin{equation}\label{non eq solid 0 order} I_\text{S.S.} = \int d^4 x \big[T^{\mu\nu} \partial_\mu X_{a\nu} + J^{i\mu} \partial_\mu \psi_a^i \big],\end{equation} 
where 
\begin{equation}\begin{split} T^{\mu\nu} & =  \tau \frac{\partial P}{\partial \tau} u^\mu u^\nu + P \eta^{\mu\nu} + \frac{\partial P}{\partial y^{ij}} \partial^\mu \psi^i \partial^\nu\psi^j,
\\ J^{i\mu} & = 2 \partial^\mu \psi^j \frac{\partial P}{\partial y^{ij}} + u^\mu \frac{\partial P}{\partial \zeta^i },
\end{split} \end{equation}
are respectively the stress-energy tensor and $Q_i$-Noether currents. As in~(\ref{solid Lagrangian 1}), the hydrodynamic pressure, $P$ is a generic function of $\tau$, $y^{ij}$, and $\zeta^i$. And the fluid four-velocity $u^\mu \equiv \tau \partial X_r^\mu /\partial \phi^0$. Notice that the equations of motion for $X_a^\mu$ and $\psi_a^i$ are just the conservation equations $\partial_\nu T^{\mu\nu}=0$ and $\partial_\mu J^{i\mu} = 0$, respectively. 

The action~(\ref{non eq solid 0 order}) as it stands represents a solid with second sound modes and no Umklapp scattering. Umklapp scattering is a process by which the lattice-momentum of the phonons is not conserved. In other words, $Q_i$ are not conserved. However, on large distance-scales, our solid should still appear homogeneous, meaning that in this course-grained picture, $Q_i$ must still represent  true symmetries of the effective theory. How can it be that $Q_i$ are symmetries of the EFT but have no corresponding conserved currents? This seems to contradict Noether's theorem. However, in non-equilibrium  EFTs, the relationship between conserved currents and symmetries is not so straight-forward. To see how this is so, suppose that we allow the action to depend on $\psi^i_a$ without derivatives. Notice that $Q_i$ act on $\psi^i_s$ for $s=1,2$ by $\psi_s^i\to \lambda^i$, for constants $\lambda^i$. Thus, since $\psi_a^i\equiv \psi_1^i -\psi_2^i$, we find that $\psi_a^i$ are invariant under $Q_i$. With these new building-blocks, our effective action becomes
\begin{equation}\begin{split}\label{ffnnoorrdd} I_\text{Umklapp} = \int d^4 x \big[T^{\mu\nu} \partial_\mu X_{a\nu} + J^{i\mu} \partial_\mu \psi_a^i + \Gamma^i \psi^i_a +\frac{i}{2} M^{ij} \psi_a^i \psi_a^j  \big], \end{split}  \end{equation}
where $\Gamma^i$ and $M^{ij}$ are functions of $\tau$, $y^{ij}$, and $\zeta^i$.  Imposing the dynamical KMS symmetries, we have
\begin{equation}  \Gamma^i = -\frac{1}{2 T_0} M^{ij} \frac{\partial \psi_r^j}{\partial\phi^0 },  \end{equation} 
where $T_0$ is the equilibrium temperature. 
 Now the equations of motion for $\psi_a^i$ are
\begin{equation} \partial_\mu J^{i\mu} = \Gamma^i.  \end{equation} 
We therefore see that the current associated with $Q_i$ is no longer conserved, as desired. Working in the large Umklapp scattering limit, the above equation simplifies to $\Gamma^i = 0$, which is solved by fixing $\partial \psi_r^i/\partial\phi^0 =0$. The equations of motion for $\psi_r^i$ give $\psi_a^i = 0$. Thus, if we integrate out $\psi_{r,a}^i$, we find that
\begin{equation}\frac{\partial \psi_r^i}{\partial\phi^0}  = 0,~~~~~\psi_a^i = 0,\end{equation} 
which are precisely the relevant IH constraints of~(\ref{IH stuff.}) necessary to remove second sound from solids. 

In summary, these second-sound-removing IH constraints arise whenever Umklapp scattering destroys the conservation of the Noether current associated with an internal translation generator. 

Curious readers may wonder how our action can have a symmetry without a corresponding Noether current. Notice that the action of $Q_i$ is to shift $\psi_r^i\to \psi_r^i +\lambda^i$, while it has no effect on $\psi_a^i$. As a result  the corresponding conserved currents furnished by Noether's theorem,  
\begin{equation}K^{i\mu} \equiv \frac{\partial I_\text{Umklapp}}{\partial (\partial_\mu \psi_r^i) } ,\end{equation} 
are conserved on-shell, namely $\partial_\mu K^{i \mu}=0$. However, since all of the terms of $I_\text{Umklapp}$ have at least one $a$-type field, all terms of $K^{i\mu}$ similarly have at least one $a$-type field. On shell, all $a$-type fields vanish, meaning that on shell, $K^{i\mu}$ must also vanish. Thus the Noether currents associated with $Q_i$ contain no physical content. 

Finally, let us return to the superfluid case. Notice that the constraint~(\ref{IH stuff.}) requires that $\psi_a^0=i\beta_0$ in order to be consistent with dynamical KMS symmetry. However, the equations of motion force all $a$-type fields to vanish, meaning that we cannot interpret the IH constraint as arising from equations of motion as we did in the case of solids. One way to remedy the situation is to instead fix $\partial_0 \psi_r^0=1$ and $\psi_a^0=0$. Such a constraint is not consistent with the dynamical KMS conditions, but it does allow us to interpret the IH constraints as arising from equations of motion when the internal shift symmetry (generated by $Q_0$) does not correspond to any conserved quantity. Since the KMS symmetries are no longer satisfied, the equilibrium state of such a system is decidedly non-thermal. Why should we need a non-thermal equilibrium state in order to remove second sound via the equations of motion for superfluids but not for solids? The reason is that if $Q_0$ is not conserved, then the equilibrium state of our system can only exist at finite density---and hence finite chemical potential (i.e. $\partial_0\psi_r^0\neq 0$)---if it is driven by some external force. But this means the equilibrium state is not thermal equilibrium. We leave the investigation of such driven systems for future work.

\section{Other worldvolumes}\label{other worldvolumes}

Thus far, we have been constructing our effective actions on the physical spacetime. While this is a valid thing to do in the classical limit, if we want a quantum theory, then we must define our EFT on a manifold other than the physical spacetime~\cite{Landry,H. Liu,H. Liu 2,H. Liu 3}. In the usual non-equilibrium coset construction of~\cite{Landry}, this manifold is the fluid worldvolume $\phi^M$ for $M=0,1,2,3$ with gauge symmetries 
\begin{equation}\begin{split}\label{the fluid symmetries} \phi^0&\to\phi^0+f(\phi^I),
\\ \phi^I&\to g^I(\phi^J),\end{split} \end{equation}
where $f$ and $g^I$ are arbitrary functions of the spatial coordinates $\phi^I$ for $I=1,2,3$. Supposing our theory is defined on the fluid worldvolume, we are interested in the effect of imposing the second-sound-removing IH constraints 
\begin{equation}\begin{split}\label{IH CMT} \frac{\partial \psi_r^{\bar\m}}{\partial\phi^0} = \delta^{\bar \m}_0,
~~~~~ \psi_a^{\bar \m}=0.  \end{split}\end{equation}
Then, just as in the classical case, we find that we may replace all instances of $\phi^{\bar\m}$ with $\psi_r^{\bar \m}$. Really, if we wanted to be completely general, we could write $\phi^{\bar \m}=\psi_r^{\bar\m}+h^{\bar \m}(\phi^I)$, for $I=1,2,3$, for some arbitrary spatially-varying function. We therefore consider $\phi^{\bar \m}=\psi_r^{\bar\m}$ to be a gauge-fixing condition equivalent to $h^{\bar\mu}=0$. With this gauge-fixing condition, we can define the `condensed-matter worldvolume' coordinates $\sigma^M$ by
\begin{equation}\sigma^M = \begin{array}{cc}
  \bigg\{ & 
    \begin{array}{cc}
      \psi_r^{\bar \m} & M =\bar\m \\
      \phi^M & M \neq \bar\m .
          \end{array}
\end{array} \end{equation} 
Thus, the gauge symmetries that $\sigma^M$ enjoy are reduced; however, whatever symmetries the fields $\psi_r^{\bar\mu}$ possess, the condensed-matter worldvolume coordinates inherit. 

To give concrete examples of what these new worldvolume symmetries look like we will give specific examples for a few condensed matter systems. For the sake of brevity, we will focus on systems for which all possible constraints of the form (\ref{IH CMT}) that can be imposed are imposed. They are as follows:
\begin{itemize}
\item {\it Superfluids:} For constant $c^0$ and arbitrary spatially-varying functions $g^i$, we have\footnote{We remind the reader that a theory constructed on this worldvolume does not describe the kind of superfluids found in nature or the laboratory. }
\begin{equation}\label{superfluid wv}\sigma^0\to\sigma^0+c^0,~~~~~\sigma^i \to g^i(\sigma^j). \end{equation}
\item {\it Solids:} For arbitrary spatially-varying function $f(\sigma^i)$ and constants $c^i$, we have
\begin{equation}\label{solid diffs} \sigma^0\to\sigma^0+f(\sigma^i),~~~~~\sigma^i \to \sigma^i+c^i. \end{equation}
Sometimes solids have the additional symmetries
\begin{equation} \sigma^i \to R^{ij}\sigma^j,\end{equation}
where $R\in SO(3)$ for isotropic solids and $R\in SO(2)$ for uniaxial crystals. 
\item {\it Supersolids:} For constants $c^M$, we have 
\begin{equation}\sigma^M\to \sigma^M+c^M.\end{equation} 
Isotropic and uniaxial supersolids also have the symmetry
\begin{equation}\sigma^i\to R^{ij} \sigma^j,\end{equation}
for $R\in SO(3)$ and $R\in SO(2)$, respectively. 
\item {\it Smectic liquid crystals in phase A:} For arbitrary spatially-varying functions $f$ and $g^A$ and constant $c^3$, we have
\begin{equation}\label{phase A wv}\begin{split}\sigma^0&\to\sigma^0+f(\sigma^i),
\\ \sigma^A &\to\sigma^A+g^A(\sigma^i),
\\ \sigma^3 &\to\sigma^3+c^3. \end{split}\end{equation}
\item {\it Smectic liquid crystals in phase B:} For arbitrary spatially-varying function $f$ and arbitrary functions $g^A$ of $\sigma^3$, and for constant $c^3$, we have
\begin{equation}\begin{split}\label{phase B wv}\sigma^0&\to\sigma^0+f(\sigma^i),
\\ \sigma^A &\to\sigma^A+g^A(\sigma^3),
\\ \sigma^3 &\to\sigma^3+c^3. \end{split}\end{equation}
\end{itemize}
Notice that all of these condensed-matter worldvolume diffeomorphism symmetries are subsets of the fluid diffeomorphism symmetries~(\ref{the fluid symmetries}).

\subsection{The solid-worldvolume coset construction}

It turns out that if we are committed to describing systems without second sound---or at least reduced second sound---we can skip over the procedure of first defining the theory on the fluid worldvolume and then imposing IH constraints.  Instead, we can define our theory directly on the condensed-matter worldvolume from the start. To demonstrate how this construction is done, we will investigate the example of an anisotropic crystalline solid with no second-sound mode using a new kind of non-equilibrium coset construction defined on the solid worldvolume. 

The only symmetries of the theory that appear in the coset are the Poincaré symmetries; in particular there are no internal translation-symmetry generators $Q_i$. Physically, we must remove the charges $Q_i$ because they are now realized as the translation gauge symmetries on the solid worldvolume coordinates~(\ref{solid diffs}). 

To keep things simple, we will work to leading order in the derivative expansion. As a result, we may construct an ordinary action with just one copy of the fields~\cite{Landry}. 
Parameterizing the most general group element by 
\begin{equation} g(\sigma) = e^{i X^\m(\sigma) P_\m} e^{i\theta^i(\sigma) J_i} e^{i \eta^i(\sigma)K_i},  \end{equation}
we find the resulting Maurer-Cartan form is 
\begin{equation}\begin{split}\label{vierbein spin} E^\mu_M &= \partial_M X^\nu {[\Lambda R]_\nu}^\m,
\\ \nabla_\m \eta^i & = (E^{-1})^M_\mu [ \Lambda^{-1}\partial_M \Lambda  ]^{0j} R^{ji},
\\ \nabla_\m\theta^{i} &=\frac{1}{2}\epsilon^{ijk}(E^{-1})^M_\m [R^{-1} \Lambda^{-1}\partial_M (\Lambda R) ]^{jk},
\end{split} \end{equation}
such that $R^{ij}=[e^{{i} \theta^{i}(\phi)J_{i}}{]^{ij}}$ and ${\Lambda^\m}_\n=[e^{{i} \eta^{i}(\phi)K_{i}}{]^\m}_\n$. We are interested in finding building-blocks that transform in a {\it manifestly} covariant fashion under~(\ref{solid diffs}).\footnote{Because rotations are spontaneously broken, we will impose no right-acting, time-independent rotation gauge symmetry on $g(\sigma)$ as was done in the fluid case in~\cite{Landry}. }

To remove boost Goldstones, impose the IH constraints $E^i_0=0$, which can be solved to give
\begin{equation} \frac{\eta^i}{\eta}\tanh \eta = -\frac{\partial_0 X^i}{\partial_0 X^t}, \end{equation}
where $\eta\equiv \sqrt{\eta^i\eta^i} $. This gives us our first building-block $E_0^t=\sqrt{-G_{00}},$ where 
\begin{equation}G_{00} = \frac{\partial X^\m}{\partial \sigma^0}\eta_{\m\n}\frac{\partial X^\n}{\partial \sigma^0}. \end{equation}
Next, impose $\epsilon^{ijk} (E^{-1})^i_j=0 $. This IH constraint tells us that
\begin{equation} (E^{-1})^i_j  = (G^{1/2})^{ij} ,\end{equation} 
where $G^{ij} = \eta^{\m\n} (e^{-1})_\m^i (e^{-1})_\n^j$ and $e_M^\m\equiv \partial_M X^\m$. We therefore can identify $G^{ij}$ as the spatial components of the inverse pull-back metric. 
Thus, the leading-order effective action is
\begin{equation}S_\text{EFT} = \int d^4 \sigma \sqrt{-G}~ P(G_{00},G^{ij}). \end{equation}
Converting to physical spacetime, we find that our action is, up to relabeling of fields, the anisotropic version of~(\ref{solid Lagrangian 2}). 

Lastly, we note that real-world solids are composed of atoms and molecules, the number of which tend to be conserved. As a result, it is often necessary to include an unbroken Goldstone associated with the $U(1)$ charge arising from particle-number conservation, that we denote by $\varphi$. Since it is an unbroken Goldstone, it enjoys the time-independent chemical shift symmetry $\varphi\to\varphi+F(\sigma^i)$. At leading order in the derivative expansion, the only symmetry-invariant building-block associated with $\varphi$ is the local chemical potential 
\begin{equation} \label{chemical potential 0} \mu = \frac{\partial \varphi}{\partial \sigma^0} .  \end{equation}
In equilibrium, we expect that $\vev{\mu}=\mu_0$ for some constant $\mu_0$. As a result, $\varphi$ must have a time-dependent equilibrium profile, namely $\vev{\varphi} = \mu_0 t$. 
Thus, the leading-order effective action including particle number conservation is
\begin{equation}\label{Solids chemical potential action 0} S_\text{EFT} = \int d^4 \sigma \sqrt{-G}~ P(G_{00},G^{ij},\mu). \end{equation}

\subsection{Classifying states of matter}

It is interesting to note that nowhere in the above coset construction did we ever need to specify the symmetry-breaking pattern. Instead, all we did was specify the global symmetry group, namely the Poincaré group, and then we specified the relevant gauge symmetries~(\ref{solid diffs}). Further, notice that the solids with and without second sound are identical at the level of SSB patterns; however, they are not identical at the level of specifying the global and gauge symmetries. In particular, at the level of the coset construction, the global symmetry group for solids with second sound is the tensor product of the Poincaré group and the internal $[U(1)]^3$ group generated by $Q_i$; the gauge group is given by~(\ref{the fluid symmetries}). On the other hand, the global symmetry group for solids without second sound is just the Poincaré group and the gauge group is given by~(\ref{solid diffs}). We therefore claim that if one wishes to be very precise, it is better to characterize states of matter according to their global and emergent gauge symmetries than by their SSB patterns. 

Finally, it is worth pointing out that this new classification in terms of emergent gauge symmetries works even in the case of zero-temperature SSB. Supposing that we have the symmetry-breaking pattern $\cG\to\cH$ at zero temperature. Then, we could equally well specify this state of matter by specifying the global symmetry group $\cG$ and then require invariance under the local right-action of $\cH$. More specifically, parameterize the most general element by
 \begin{equation}\label{zero T parameterization 2}g[\pi,\epsilon,x) = e^{i x^\mu \bar P_\mu} e^{i\pi^\alpha(x)\tau_\alpha} e^{i \epsilon^A(x)T_A}. \end{equation} 
Then if we require invariance under the gauge transformation 
\begin{equation}g[\pi,\epsilon,x) \to g[\pi,\epsilon,x) \cdot h(x), \end{equation}
for generic $h(x)\in \cH_0$, this forces all unbroken Goldstones $\epsilon^A(x)$ to be pure gauge. As a result, they cannot possibly appear in the invariant building-blocks, so we will construct the same effective action with this method as we would with the usual coset parameterized by~(\ref{zero T parameterization}).

\section{Solids and smectics with dissipation}

We begin by constructing the effective action for solids without second sound to leading order in dissipation; then we will show how to modify it to account for smectics in phases A and B. The construction of such an action requires doubled field content. We could use the coset construction to formulate this higher-order action, but we find it convenient to use a different method that makes the constitutive relations more apparent. 

We will work exclusively in the classical limit, allowing us to formulate our action on the physical spacetime coordinate $x^\mu\equiv X_r^\mu$. As a result, the solid worldvolume coordinates become dynamical fields. Our field content is now $\sigma^M(x)$, $\varphi_r(x)$, $X_a^\mu(x)$, and $\varphi_a(x)$. 

To construct the effective action, it is helpful to first identify the symmetry covariant building-blocks. The retarded building-blocks are as follows. Let $K_\mu^M(x) \equiv \partial_\mu\sigma^M(x)$. Then the local inverse-temperature four-vector field is given by 
\begin{equation}\beta^\mu(x) = \beta_0 (K^{-1})_0^\mu , \end{equation}
where $\beta_0$ is the equilibrium inverse temperature. 
This field encodes all information about the temperature and local rest-frame of the solid volume elements. It is often helpful to decompose this object into its magnitude and direction by $\beta^\mu = \beta u^\mu$, where
\begin{equation}\beta = \sqrt{-\beta^2},~~~~~~~~~~u^\mu = \frac{\beta^\mu}{\beta}. \end{equation}
Next, there are the solid basis vectors and (inverse) solid metric given respectively by
\begin{equation} e_\mu^i(x) = K_\mu ^i,~~~~~~~~~~ \gamma^{ij}(x) = e_\mu^i  \eta^{\mu\nu} e_\nu^j.  \end{equation}
Note that $i,j,k,l = 1,2,3$ indicate spatial solid coordinate indices as opposed to physical-space indices. We also have the chemical potential $\mu$ given by~\eqref{chemical potential 0} as a building-block. 
In addition to the retarded building-blocks, we have the advanced covariant building-blocks 
\begin{equation} G_{a\mu\nu} = \partial_\mu X_{a\nu} +\partial_\nu X_{a\mu},~~~~~ B_{a\mu} = \partial_\mu \varphi_a .  \end{equation}

The leading order action is constructed with $\beta$, $\mu$, $u^\mu$, $\gamma^{ij}$, $G_{a\mu\nu}$, and $B_{a\mu}$ without any additional derivatives. Further, because advanced fields count at higher-order in the derivative expansion, the leading terms may only contain one factor of $G_{a\mu\nu}$ and $B_{a\mu}$. Thus, the leading-order Lagrangian is
\begin{equation}\cL_1 = \frac{1}{2} T^{\mu\nu}_0 G_{a\mu\nu} + J_0^\mu B_{a\mu} , \end{equation}
where $T_0^{\mu\nu}$ is some symmetric tensor and $J_0^\mu$ is some four-vector built from $\beta$, $\mu$, $u^\mu$, and  $\gamma^{ij}$ and have the interpretation of the stress-energy tensor and particle number current, respectively. Notice that the equations of motion for $X_a^\mu$ and $\varphi_a$ are, respectively $\partial_\nu T^{\mu\nu}_0=0$ and $\partial_\mu J_0^\mu=0$. The most general form this leading-order stress-energy tensor can take is
\begin{equation} T_0^{\mu\nu} = \epsilon_0 u^\mu u^\nu + p_0 \Delta^{\mu\nu} + r_{ij} \Delta^{ij \mu\nu}  ,\end{equation}
where $\Delta^{\mu\nu} = g^{\mu\nu} +u^\mu u^\nu$ and $\Delta^{ij}_{ \mu\nu} = \frac{1}{2} (e^i_\mu e^j_\n+e^j_\mu e^i_\nu)$. And the leading-order particle number current is
\begin{equation} J_0^\mu=n_0 u^\mu. \end{equation}
We take $\epsilon_0$, $p_0$, $r_{ij}$, and $n_0$ to be generic functions of $\beta$, $\mu$, and $\gamma^{ij}$. 
For isotropic solids, the sum over indices $i,j$ must be performed in a rotationally-invariant manner, but for generic solids, the $i,j$ indices on $r_{ij}$ and $\gamma^{ij}$ are purely for the purposes of bookkeeping and need not transform in any particular way. 

At leading order, the dynamical KMS symmetries allow us to write the action in factorized form as $\int d^4 x\cL_1 = S_\text{EFT}[X_1] -S_\text{EFT}[X_2] + \cO(a^3)$, where $S_\text{EFT}$ is given in~\eqref{Solids chemical potential action 0}. As a result, we have the relations among $\epsilon_0$, $p_0$ and $r_{ij}$ given by 
\begin{equation} \label{KMS leading order}p_0\equiv P(\beta,\mu,\gamma^{ij}),~~~~~ \epsilon_0 + p_0 =  -\beta \frac{\partial p_0}{\partial \beta }+\mu \frac{\partial p_0}{\partial \mu},~~~~~ r_{ij} = \frac{\partial p_0}{\partial \gamma^{ij}},~~~~~n_0=\frac{\partial p_0}{\partial \mu} . \end{equation} 

The next-to-leading-order (NLO) terms in the Lagrangian give rise to dissipation. We have
\begin{equation}\cL_2 = \frac{1}{2} T_1^{\mu\nu} + J_1^\mu B_{a\mu} +\frac{i}{4} W_0^{\mu\nu,\alpha\beta} G_{a\mu\nu} G_{a\alpha\beta}+i Z_0^{\mu\nu} B_{a\mu}B_{a\nu} . \end{equation}
Physically, $T_1^{\mu\nu}$ and $J_1^\mu$ are, respectively the NLO contributions to the stress-energy tensor and the particle number current. By contrast, $W_0^{\mu\nu,\alpha\beta}$ and $Z_0^{\mu\nu}$ encode information about statistical fluctuations. The explicit forms of these terms are potentially quite complicated. Fortunately, we can employ the generalized Landau frame~\cite{H. Liu} to simplify matters. In particular, we have
\begin{equation}T_1^{\mu\nu} = - \eta_{ijkl} \Delta^{ij \mu\nu} \Delta^{kl \alpha \beta} \partial_\alpha u_\beta,~~~~~~~~~~ W_0 ^{\mu\nu,\alpha\beta} = \beta^{-1}\eta_{ijkl} \Delta^{ij \mu\nu} \Delta^{kl \alpha \beta}  , \end{equation}
and
\begin{equation} J_1^\mu = -\sigma_{ij} \Delta^{ij \mu\nu} \beta^{-1} \partial_\nu (\beta \mu),~~~~~~~~~~Z_0^{\mu\nu} = \beta^{-1} \sigma_{ij} \Delta^{ij \mu\nu}, \end{equation}
where the forms of $W_0^{\mu\nu,\alpha\beta}$ and $Z_0^{\mu\nu}$ are determined by the classical dynamical KMS symmetries.\footnote{In general, the field redefinitions required to arrive at the generalized Landau frame do not respect the dynamical KMS symmetries. Conveniently, however, the dynamical KMS symmetries hold for the NLO Lagrangian in generalized Landau frame. }
We interpret $\eta_{ijkl}$ as the viscosity tensor and $\sigma_{ij}$ as the charge conductivity tensor, which is related to the thermal conductivity tensor $\kappa_{ij}$ by
\begin{equation} \kappa_{ij} = \bigg(\frac{\epsilon_0+p_0}{n_0}\bigg)^2 \beta \sigma_{ij}. \end{equation} 
These tensors enjoy various symmetries among their indices, namely
\begin{equation}\eta_{ijkl} = \eta_{jikl}=\eta_{ijlk}=\eta_{klij},~~~~~\sigma_{ij}=\sigma_{ji}. \end{equation}

Putting it all together, the full Lagrangian to leading order in dissipation is $\cL=\cL_1+\cL_2$, or explicitly, 
\begin{equation}\begin{split} \cL =\frac{1}{2}\bigg [\epsilon_0 u^\mu u^\nu + p_0 \Delta^{\mu\nu} + r_{ij} \Delta^{ij \mu\nu} - \eta_{ijkl} \Delta^{ij \mu\nu} \Delta^{kl \alpha \beta} \bigg( \partial_\alpha u_\beta-\frac{i}{4\beta} G_{a\alpha\beta} \bigg)\bigg]G_{a\mu\nu} \\
+ \big [n_0 u^\mu -\sigma_{ij} \Delta^{ij \mu\nu} \beta^{-1} \big( \partial_\nu (\beta \mu) - iB_{a\nu} \big)\big] B_{a\mu} .
 \end{split} \end{equation}

Lastly, it is worth noting that the effective actions for smectic liquid crystals in phases A and B can be obtained quite easily from the above action. In fact, they are special cases of the above. Consider the symmetries~\eqref{phase A wv} and~\eqref{phase B wv}. Notice that they contain strictly more symmetries than the solid worldvolume. In this way smectic liquid crystals can be seen as symmetry enhanced solids. To obtain the action for smectic A liquid crystals, we restrict the kind of dependence the action can have on $\gamma^{ij}$. In particular, the action may only depend on $\gamma^{33}$ and the transport coefficients $\eta_{ijkl}$ and $\sigma_{ij}$ must be rotationally symmetric about the $3$ direction. Explicitly, for any rotation
\begin{equation}
R^{ij}(\theta) = 
\begin{pmatrix}
\cos\theta & -\sin\theta & 0\\
\sin\theta & \cos\theta & 0\\
0 & 0 & 1
\end{pmatrix}^{ij},
\end{equation}
we have that
\begin{equation}\eta_{ijkl} = R^{ii'} R^{jj'} R^{kk'} R^{ll'} \eta_{i'j'k'l'} ,~~~~~\sigma_{ij} =R^{ii'} R^{jj'} \sigma_{i'j'} . \end{equation}
The action for smectic B liquid crystals has the same building-blocks as that of phase A except its dependence on $\gamma^{ij}$ may also include
\begin{equation} b = \det \gamma,~~~~~ b_1 = \gamma^{11}\gamma^{33}- (\gamma^{13})^2,~~~~~b_2 = \gamma^{22} \gamma^{33} - (\gamma^{23})^2,   \end{equation}
and the viscosity tensor need not have any rotational symmetry about the 3 direction. 
Comparing with~\cite{H. Liu}, we see that fluids and smectic liquid crystals can be viewed as highly symmetric solids at the level of the non-equilibrium effective action.

\subsection{Comparing with previous results}

Recent papers~\cite{Armas:2020bmo,Armas:2019sbe} have constructed effective field theories for relativistic solids by formulating constitutive relations for conserved quantities. The constitutive relations that these authors arrive at bear great resemblance to those presented here, but with some differences. In particular, in our formulation, the solid basis vectors $e_\mu^i$ are automatically orthogonal to the fluid velocity $u^\mu$. That is $u^\mu e_\mu^i\equiv 0 $ off shell. This orthogonality arises because we have removed second sound; if second sound were present then the fluid degrees of freedom could flow freely relative to the solid degrees of freedom, thereby permitting $u^\mu e_\mu^i\neq 0$. Alternatively even with second sound present, we could abstain from imposing the $\psi_s$-removing IH constraints and find that, on the equations of $u^\mu e_\mu^i$ is non-zero but decays to zero exponentially fast. In particular if $\Gamma^i$ from~\eqref{ffnnoorrdd} is sufficiently large, then the exponential decay will take place on shorter time scales than the UV cutoff of the EFT; if $\Gamma^i$ is sufficiently small, then such decays occur on time scales longer than the UV cutoff. 

As a result, the results presented in this paper agree with those of~\cite{Armas:2020bmo,Armas:2019sbe} so long as we augment their equations with additional equations either forcing $u^\mu e_\mu^i\equiv 0$ or equations dictating the (non)-conservation of the lattice momentum currents $J^{i\mu}$.

\section{Summary} 

In this paper, we identified the key ingredient---from the perspective of non-equilibrium effective field theory---that distinguishes condensed matter systems with and without second-sound modes. In particular, we found that an IH constraint can be imposed to remove second-sound modes at the level of the non-equilibrium coset construction. Unlike other IH constraints, however, the existence of these new constraints are not derivable from the usual algebraic relations involving commutators of various symmetry generators. The only thing they have in common with the usual IH constraints is that they allow the removal of one set of fields in favor of another. 

After identifying these new IH constraints, we then demonstrated how they can be applied to various states of matter including superfluids, isotropic solids, uniaxial crystals, supersolids, and smectic liquid crystals in phases A and B. In all of these examples, it was possible (though in the cases of superfluids and supersolids, not necessarily physically reasonable) to remove the second-sound modes with IH constraints, but some states of matter also admit the unusual possibility of partial removal of second-sound modes. In particular, by imposing some but not all of the possible IH constraints, we found that it is possible to have second-sound modes that can propagate in some directions but not others. It would be fascinating to see if any such states of matter exist in nature.

Additionally, we demonstrated that if we are committed to describing condensed matter systems without second sound---or at least partially removed second sound---then we can construct our theory directly on a new condensed matter worldvolume, as opposed to the usual fluid worldvolume, thus circumventing the need for the new IH constraints. The difference between the fluid worldvolume and other condensed-matter worldvolumes has to do with the diffeomorphism gauge symmetries that exist in each. In particular, the fluid worldvolume has symmetries given by~(\ref{the fluid symmetries}), whereas condensed matter worldvolumes have reduced gauge symmetries; see (\ref{superfluid wv}-\ref{phase B wv}) for examples. As a concrete demonstration, we  formulated the leading-order effective action for anisotropic solids using the coset construction defined on the solid worldvolume. 

We noted that instead of using Landau's classification of states of matter in terms of SSB patterns, we can be more precise if we merely specify the global and emergent gauge symmetries without making reference to SSB at all. We hope that more exotic states of matter for which Landau's system is inadequate can be understood in terms of this new classification method. 

Further, we found that the physical origins of the IH constraints responsible for removing second sound are directly related to Umklapp scattering. Thus, it is appropriate to impose such IH constraints for EFTs of certain solids, but not for superfluids since superfluids do not exhibit Umklapp scattering. 

Finally, we constructed the effective actions for solids and smectic liquid crystals in phases A and B to leading order in the derivative expansion. We found that in the absence of second sound, fluids and smectic liquid crystals can be viewed as highly symmetric solids. 

\bigskip

\noindent {\bf Acknowledgments:} I would like to thank Alberto Nicolis and Lam Hui for their wonderful mentorship and Matteo Baggioli for posing insightful questions that inspired this project. This work was partially supported by the US Department of Energy grant DE-SC011941.

\appendix

\end{document}